\documentclass[twocolumn,amssymb,showpacs, amsmath,nobibnotes,nofootinbib,aps,prb]{revtex4}
\usepackage[dvips]{graphicx}
\begin{document}

\title{Field-Induced Magnetization Steps in Intermetallic Compounds and Manganese
Oxides: The Martensitic Scenario}
\author{V. Hardy$^{1,2}$, S. Majumdar$^{1}$, S. Crowe$^{1}$, M. R. Lees$^{1}$, D. Mc
K. Paul$^{1}$, L. Herv\'{e}$^{2}$, A. Maignan$^{2}$, S. H\'{e}bert$^{2}$, C.
Martin$^{2}$, C. Yaicle$^{2}$, M. Hervieu$^{2}$ and B. Raveau$^{2}$}
\affiliation{(1) Department of Physics, University of Warwick, Coventry, CV4 7AL, United
Kingdom.\\
(2) Laboratoire CRISMAT, UMR 6508, Boulevard du Mar\'{e}chal Juin,\\
14050 Caen, France. }

\begin{abstract}
Field-induced magnetization jumps with similar characteristics are observed at low temperature
for the intermetallic germanide Gd$_{5}$Ge$_{4}$ and the mixed-valent
manganite Pr$_{0.6}$Ca$_{0.4}$Mn$_{0.96}$Ga$_{0.04}$O$_{3}$. We report that
the field location -and even the existence- of these jumps 
depends critically on the magnetic field sweep rate used to record the data. It is
proposed that, for both compounds, the martensitic character of their
antiferromagnetic-to-ferromagnetic transitions is at the origin of the
magnetization steps.
\end{abstract}
\maketitle
The Gd$_{5}$(Si$_{x}$Ge$_{1-x}$)$_{4}$ pseudobinary system has attracted a
growing interest in recent years\cite
{PE97,MO98,MO00,ST00,LE01a,LE01b,PE02,LE02,SO03} owing to the wealth of
interesting physical properties it displays including a giant magnetocaloric effect\cite
{PE97} and a colossal magnetostriction.\cite{MO98} These striking phenomena
 are related to a strong interplay between the magnetic and the
structural features in this system. These compounds have a layered structure
made up of sub-nanometric slabs connected via covalent-like bonds.\cite{PE02}
The degree of interslab connectivity not only depends on $x$ but also on the
magnetic state. For instance, with $x=0$, the slabs are completely
interconnected in the ferromagnetic (FM) state, whereas all the bonds are
broken in both the antiferromagnetic (AF) and paramagnetic (P) states.\cite
{LE01b}

Recently, Levin {\it et al.}\cite{LE02} have reported an intriguing
phenomenon for Gd$_{5}$Ge$_{4}$. After zero-field-cooling (ZFC), the
field-increasing branch of $M(H)$ curves recorded at low-$T$ exhibit an
extremely sharp, irreversible magnetization step. It has been proposed that
this behavior is related to the strongly anisotropic exchange interactions
present in this material. One should note, however, that the observation
of such sharp steps in polycrystalline samples is quite unusual for
conventional metamagnetic transitions.

\begin{figure}[t]
\centering
\label{fig1}
\includegraphics[width = 8.4 cm]{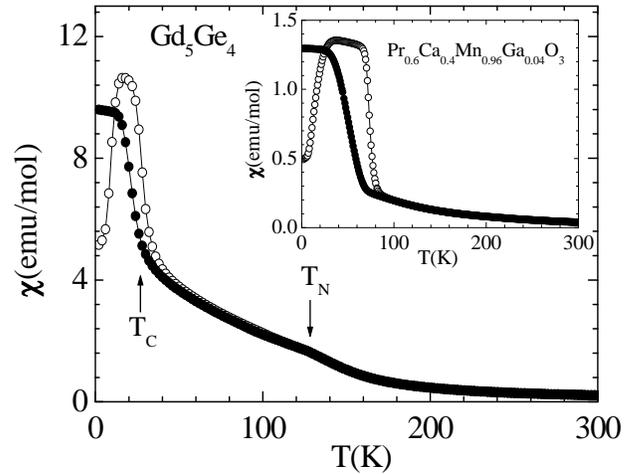}
\caption{$dc$ susceptibility curves recorded for Gd$_{5}$Ge$_{4}$ (main panel)
and Pr$_{0.6}$Ca$_{0.4}$Mn$_{0.96}$Ga$_{0.04}$O$_{3}$ (inset) in a field of
1.2 T. The open and closed symbols correspond to the zero-field cooled and
field cooled cooling modes respectively. Arrows and labels on the main
panel denote the N\'{e}el temperature ($T_{N}\sim 127$ K) and Curie
temperature ($T_{C}\sim 25$ K) of Gd$_{5}$Ge$_{4}$}
\end{figure}      

Interestingly, similar magnetization steps were recently observed for
mixed-valent manganese oxides with the general formula Pr$_{1-x}$Ca$_{x}$Mn$_{1-y}$%
M$_{y}$O$_{3}$ (with $x$ $\sim $ 0.5, $y$ $\sim $ 0.05, and where M is a
cation used to destabilize the Mn-sublattice).\cite{HE02,MAI02,MAH02} The
Mn-site substitutions weaken the robust CE-type AF ordering of the parent
compound Pr$_{1-x}$Ca$_{x}$MnO$_{3}$ ( $x$ $\sim $ 0.5), and favor the
development of a phase separation between FM and AF domains. Owing to the
collective orbital ordering (OO) accompanying the AF spin ordering, the unit
cell of the AF phase is strongly distorted with respect to that of the FM
phase. Therefore, as a magnetic field is applied, competition develops
between the magnetic energy promoting the development of the FM phase and
the elastic energy associated with the strains created at the AF/FM
interfaces, which tends to block the transformation.\cite{HA03} The
martensitic nature of this transformation has led us to propose that the
magnetization step corresponds to a burst-like growth of the FM component
when the driving force overcomes the energy barriers associated with the
strains. Remarkably, it turns out that Gd$_{5}$Ge$_{4}$ is also a system in
which FM and AF domains can co-exist, and the transformation between these
two phases has a pronounced martensitic character.\cite{LE01b,LE02} This is
due to the collective shear movement of the slabs at the AF/FM transition
which produces a considerable distortion of the unit cell. For the closely
related Gd$_{5}$(Si$_{0.1}$Ge$_{0.9}$)$_{4}$ compound, Morellon {\it et al.} 
\cite{MO00} reported that the cell parameter $a$ decreases by 1.6\%, while $%
b $ and $c$ increase by 0.7 and 0.3\%, respectively, at the AF/FM transition
($Pnma$ \ space group for both phases).

This set of features has prompted us to undertake a precise comparative
study between Gd$_{5}$Ge$_{4}$ and a Mn-site substituted manganite, with a
particular focus on the influence of the magnetic field sweep rate on the
field-induced transformations. Our goal was to further investigate the
similarity of the magnetization steps in these two systems and the relevance
of the martensitic scenario to both of them.

\begin{figure}[t]
\centering
\label{fig2}
\includegraphics[width = 8.4 cm]{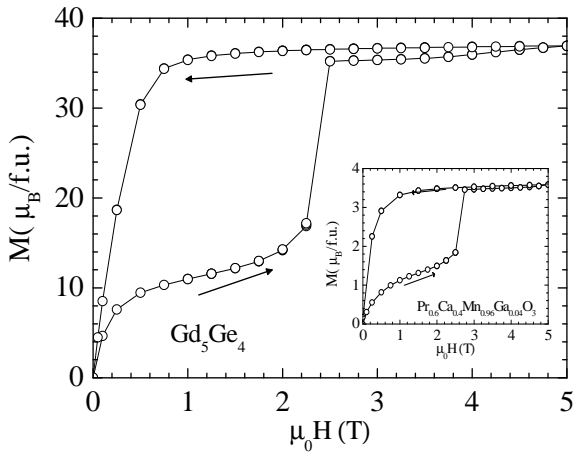}
\caption{Magnetic hysteresis loops recorded with a SQUID magnetometer after
zero-field cooling: at 2 K for Gd$_{5}$Ge$_{4}$ (main panel), and at 3.25 K
for  Pr$_{0.6}$Ca$_{0.4}$Mn$_{0.96}$Ga$_{0.04}$O$_{3}$ (inset)}
\end{figure}   

The Gd$_{5}$Ge$_{4}$ sample was prepared by arc melting a stoichiometric
mixture of 99.9 wt\% pure Gd and 99.99 wt\% pure Ge. The synthesis was
carried out under a high-purity argon atmosphere, turning the sample several
times to ensure a good homogeneity. The Gd/Ge ratio was checked by EDS
(energy dispersive spectroscopy) to be equal to the nominal composition to
within the accuracy of this technique. No impurities were detected by x-ray
powder diffraction, which showed that the system has an orthorhombic structure at
room temperature ($Pnma$ $\ $space group) with lattice parameters [$%
a=7.68(1) $ \AA $,$ $b=$ $14.80(1)$ \AA\ and $c=7.77(1)$ \AA ] in line with
the literature.\cite{LE01b} The manganite chosen to compare with Gd$_{5}$Ge$%
_{4}$ is Pr$_{0.6}$Ca$_{0.4}$Mn$_{0.96}$Ga$_{0.04}$O$_{3}$, hereafter denoted
as [PrCa40]Ga4\%. This compound exhibits a phase-separation similar to that
of Gd$_{5}$Ge$_{4}$, and it is less sensitive to training effects than most
of the manganites.\cite{MAI02} A [PrCa40]Ga4\% ceramic sample was
synthesized by solid-state reaction according to a process described
previously.\cite{HE02} It also has a room-temperature $Pnma$ \ orthorhombic structure
with $a=5.4293(3)$ \AA $,$ $b=$ $7.6443(4)$ \AA\ and $c=5.4097(3)$ \AA .
Magnetic measurements were carried out using a Superconducting Quantum
Interference Device (SQUID) magnetometer and a Vibrating Sample Magnetometer
(VSM). All the $M(H)$ curves have been recorded after being ZFC from the
paramagnetic state at 300 K.

The main panel of Fig. 1 shows the ZFC (Zero Field Cooled) and the FCC (Field Cooled
Cooling) $dc$ magnetic susceptibility ($\chi$) curves as a function of temperature for Gd$_{5}$Ge$_{4}$. 
These curves were recorded in a field of 1.2 T to be comparable with those of Ref. 8. The
inset shows the same data sets for [PrCa40]Ga4\%. The $\chi (T)$ curves of
Gd$_{5}$Ge$_{4}$ exhibit the same general features as those reported by
Levin {\it et al.}\cite{LE02}: (i) a kink at $T_{N}=127$ K, and (ii) an
increase of $\chi $ at low-$T$, which is associated with the onset of a FM\
ordering. There is a pronounced hysteresis at low-$T$ that points to the
first-order character of this ferromagnetic transition, the inflection point
on the ZFC and FCC curves being at $\sim 28$ and $\sim 21$ K, respectively.
As in Ref. 8, one can also observe that the maximum value of $\chi (T)$ is
larger for the ZFC\ data than for the FCC data, and that there is a steep
rise of the ZFC $\chi (T)$ on the low-$T$ side of the peak. It should be
noted that this last feature is quite unusual for standard ferromagnets in
a field as large as 1.2 T. It suggests that the onset of the FM phase is
hindered when zero-field cooling this compound down to very low temperature.
In addition, note that the difference between the values of the
magnetization for the ZFC and FCC curves around 20 K may be related to the
large magnetostriction present in this material. A\ Curie-Weiss fit of the
paramagnetic regime for $T>240$ K gives $\theta _{CW}=(115.2\pm 0.5)$ K and $%
\mu _{eff}=(7.85$ $\pm 0.01)$ $\mu _{B}$ / Gd, the latter value being close
to the theoretical expectation ($\mu _{eff}=7.94$ $\mu _{B}$ / Gd). This set
of parameters is also consistent with the previous study of Gd$_{5}$Ge$_{4}$
which reported $\theta _{CW}\sim 94$ K and $\mu _{eff}\sim 7.45$ $\mu _{B}$
/ Gd.\cite{LE02} A closer look at the data of Fig. 1 reveals some
differences from the sample studied by Levin {\it et al}.\cite{LE02}, in
particular a $\chi (T\rightarrow 0)$ value of the ZFC curve that is larger
by $\sim 25$ \% in our case.

\begin{figure}[t]
\centering
\label{fig3}
\includegraphics[width = 8.4 cm]{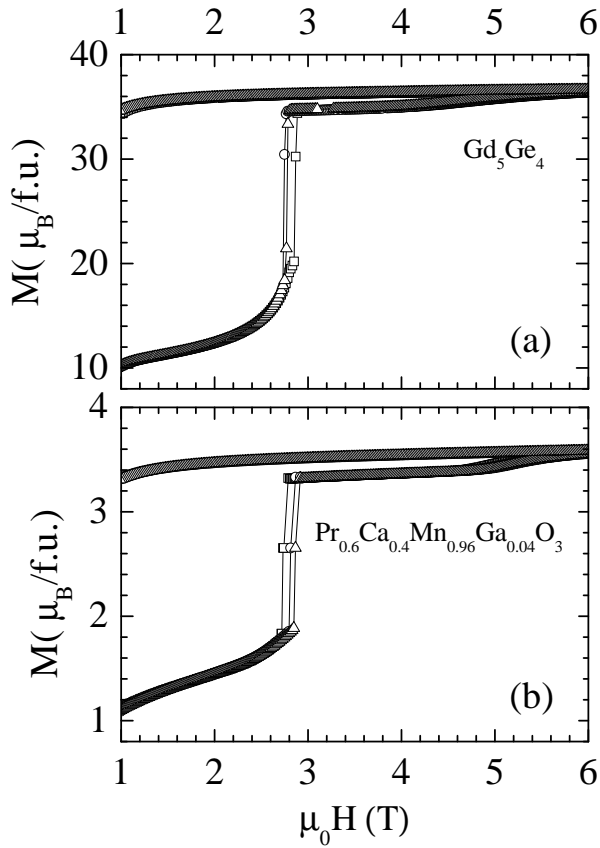}
\caption{Enlargements of three successive hysteresis loops recorded with a
VSM, after zero-field cooling in each case: (a) Gd$_{5}$Ge$_{4}$ at 2 K; (b)
Pr$_{0.6}$Ca$_{0.4}$Mn$_{0.96}$Ga$_{0.04}$O$_{3}$ at 3.25 K. The magnetic
field sweep rate is 1 T/min.}
\end{figure}

The inset of Fig. 1 shows that the $\chi (T)$ curves of [PrCa40]Ga4\%
exhibit low-$T$ features that are remarkably similar to those found in Gd$%
_{5}$Ge$_{4}$. This behavior was attributed to the appearance of a FM
component in this manganite, while electron microscopy demonstrated the
persistence, at low-$T$, of a short-range OO associated with the CE-type AF
phase.\cite{MAI02} The similarity between the $\chi (T)$ curves of Gd$_{5}$Ge%
$_{4}$ and [PrCa40]Ga4\% in the low-$T$ regime is consistent with the
existence, in both systems, of related ground states based on phase
separation between AF and FM domains.

The main panel of Fig. 2 shows a $M(H)$ curve recorded for Gd$_{5}$Ge$_{4}$
at 2 K after ZFC. For the field-increasing branch, there is a dramatic step in the 
magnetization between 2.25 and 2.50 T. This jump is followed by a
plateau, then a smooth tail, before finally reaching $M_{sat}=36.6$ $\mu
_{B} $ / f.u. at 5 T. This saturation value corresponds to 7.3 $\mu _{B}$ /
Gd, in good agreement with Ref. 8. The field-decreasing branch is almost
flat down to $\sim 1$ T before going to zero. Increasing the field once
again produces a curve superimposed on the reverse leg of the first loop,
demonstrating the complete irreversibility of the transformation at this
temperature. The overall behavior displayed in the main panel of Fig. 2 is
in line with the features reported by Levin et {\it al.}.\cite{LE02} The
difference, however, is that our sample exhibits a sizeable FM component, as
already suggested by the ZFC value of $\chi (T\rightarrow 0)$. In Fig. 2,
this is clearly revealed by the shape of the virgin magnetization curve at
low fields. It is worth noting that the behavior of the sample shown in  Fig. 2 is closer to that
described in  Ref. 8 when it is cooled in a field of 1.2 T to
assist the onset of the FM component. We suggest that the two samples
may differ on a  microstructural or nanostructural level (for example,
grain size or local defects), and this in turn may  influence the ability of the samples to accommodate
the strains associated with the martensitic (FM) phase.\cite{NIS,MAR,PO01}

The inset of Fig. 2 shows a $M(H)$ curve recorded for [PrCa40]Ga4\% at 3.25
K, after ZFC. This sample  also exhibits a magnetization jump and
all the features found for Gd$_{5}$Ge$_{4}$. For both compounds, the
location of the magnetization steps in the $M(H)$ curves depends on the
temperature. In order to obtain comparable data for each system, all the $M(H)$ loops
recorded hereafter were recorded at 2 K for Gd$_{5}$Ge$_{4}$ and at 3.25 K for
[PrCa40]Ga4\%.

In manganites, one of the features supporting a martensitic scenario rather
than standard metamagnetism was the influence of the field spacing used to
record the $M(H)$ curves by SQUID magnetometry.\cite{HA03} It was found that
smaller field increments can delay the magnetic instability, pushing the
steps to higher field values. With SQUID measurements such a field-spacing
effect can be related to the average magnetic field sweep rate. In
the present study, we have used a VSM which is more suited to properly
address this issue, since the data can be recorded while ramping the field.
First of all, one must check that the compounds are not too sensitive to
training effects (i.e., the shift of the step fields between successive ZFC hysteresis
loops).

Figure 3 shows enlargements of three successive ZFC $M(H)$ loops recorded on
Gd$_{5}$Ge$_{4}$ and [PrCa40]Ga4\%. One can observe small variations from
run to run in the value of the step field, which is found to be $2.78\pm
0.08 $ T in both cases. Additional ZFC loops recorded in the case of Gd$_{5}$%
Ge$_{4}$ showed that the step field always lies within this range. The
difference between these data and the measurements made using a SQUID magnetometer (shown in Fig. 2)
 will be discussed below. One should note that the scatter displayed in Fig. 3 -even though it is small-
indicates that the step field of both compounds does not correspond to a
well defined critical field, whereas such an history dependence is
consistent with a martensitic picture. In other respects, these training
effects for Gd$_{5}$Ge$_{4}$ and [PrCa40]Ga4\% remain small enough to allow
a reliable investigation of the influence of the magnetic field sweep rate.

\begin{figure}[t]
\centering
\label{fig4}
\includegraphics[width = 8.4 cm]{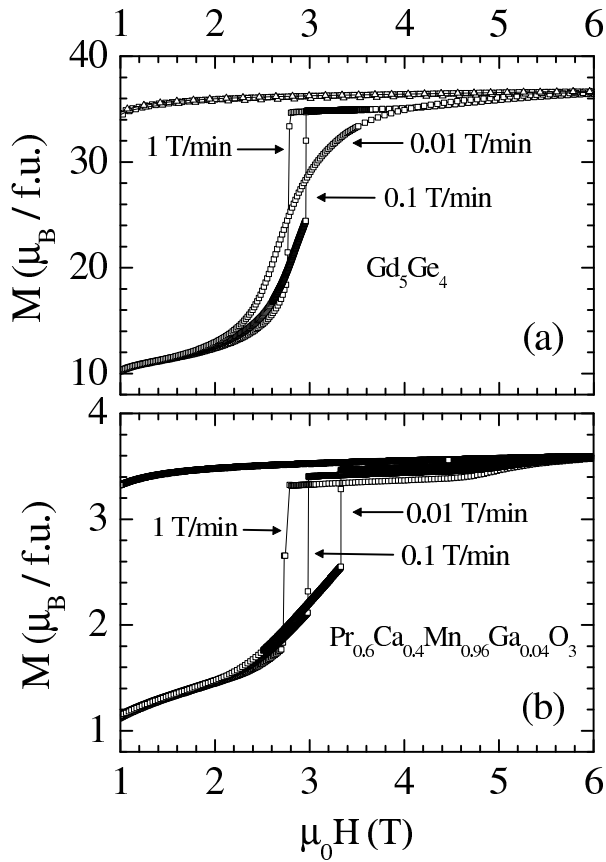}
\caption{Enlargements of hysteresis loops recorded with a VSM using different
magnetic field sweep rates, after zero-field cooling in each case: (a) Gd$%
_{5}$Ge$_{4}$ at 2 K; (b) Pr$_{0.6}$Ca$_{0.4}$Mn$_{0.96}$Ga$_{0.04}$O$_{3}$
at 3.25 K.}
\end{figure}

Figure 4 shows enlargements of $M(H)$ loops recorded on Gd$_{5}$Ge$_{4}$ and
[PrCa40]Ga4\% with three magnetic field sweep rates $dH/dt=\dot{H}$,
 ranging over two orders of magnitude. {\it In both systems, the
magnetization step is found to be profoundly affected by the value of }
$\dot{H}$. Once again, this confirms that these step fields
cannot be regarded as true critical fields for metamagnetic transitions. It
appears that the influence of $\dot{H}$ is similar in both
systems; as $\dot{H}$ is reduced, the smooth upturn of $M(H)$
starts at lower fields, whereas the step in the magnetization is pushed to a higher field. 
For Gd$_{5}$Ge$_{4}$, this effect is so pronounced  that there is no longer a step
for $\dot{H}$ $=0.01$ T/min. A similar disappearance of the
step as $\dot{H}$ is decreased was also observed for
[PrCa40]Ga4\% at 3.5 K for 0.1 T/min (not shown). Such a huge impact of the
magnetic field sweep rate is a feature that can be accounted for in a
martensitic interpretation of the magnetization steps. Indeed, for
isothermal martensitic transformations, it is known that the rate of
variation of the driving force (here the magnetic field) can affect the
development of the transformation. For instance, P\'{e}rez-Reche {\it et al.}
\cite{PER01} have recently reported a significant influence of the cooling
rate $dT/dt$ on the temperatures of the peaks displayed on the acoustic
emission spectra of Cu$_{68.4}$Al$_{27.8}$Ni$_{3.8}$. In our case, the
effect is found to be more systematic and pronounced. We suggest that using a
smaller $\dot{H}$ can facilitate the progressive
accommodation of the martensitic strains, resulting in an upward shift of
the step field and even its disappearance.

Let us now return to the comparison between the SQUID magnetometer and VSM data.
According to the systematic influence of $\dot{H}$ shown by
the VSM data, the locations of the step field in the SQUID data point to an
effective sweep rate close to or even slightly larger than 1 T/min. For the
SQUID measurements of Fig. 2, the average sweep rate (including the pause
and measurements at each field) is $\sim 0.05$ T/min, while the transitory
sweep rate when charging the magnet is $\sim 2$ T/min. Our results suggest
that (i) the average sweep rate is not a relevant parameter to characterize
the dynamics when the magnetic field has to be stabilized prior to each
measurement; (ii) using a fast sweep rate for the field installations can
play an important role in  determining the response of these martensitic systems.

The present paper demonstrates that the magnetization steps recently reported
for Gd$_{5}$Ge$_{4}$ have features very similar to those found in manganites
like [PrCa40]Ga4\%, including a huge influence of the magnetic field sweep
rate on the field-induced transformations. Such a feature is inconsistent
with a standard metamagnetic transition whereas it can be qualitatively
accounted for within a martensitic scenario. Although they belong to
completely different classes of materials, both Gd$_{5}$Ge$_{4}$ and
[PrCa40]Ga4\% turn out to be phase-separated systems, in which FM and AF
domains having very different unit cells can co-exist. Therefore, for both
systems, the field-induced AF-to-FM transition at low-$T$ must be regarded
as a martensitic transformation. Such transformations are well known to be
discontinuous, and they can show burstlike effects. Accordingly, we propose
that the similarity of the low-$T$ properties found in Gd$_{5}$Ge$_{4}$ and
[PrCa40]Ga4\% is not coincidental, and that the magnetization steps are
manifestations of the martensitic nature of the transformation in both
systems. In this scenario, the magnetization jump corresponds to a burstlike
growth of the FM component within an essentially AF matrix.
\par
We acknowledge the financial support of the EPSRC (UK) for this project.


\begin{thebibliography}{99}
\bibitem{PE97}  V. K. Pecharsky and K. A. Gschneidner, Phys. Rev. Lett. {\bf
78}, 4494 (1997).

\bibitem{MO98}  L. Morellon {\it et al.}, Phys. Rev. B {\bf 58}, 14721
(1998).

\bibitem{MO00}  L. Morellon {\it et al.}, Phys. Rev. B {\bf 62}, 1022
(2000).

\bibitem{ST00}  J. Stankiewicz {\it et al.}, Phys. Rev. B {\bf 61}, 12651
(2000).

\bibitem{LE01a}  E. M. Levin {\it et al.}, Phys. Rev. B {\bf 63}, 064426
(2001).

\bibitem{LE01b}  E. M. Levin {\it et al.}, Phys. Rev. B {\bf 64}, 235103
(2001).

\bibitem{PE02}  A. O. Pecharsky et al., J. Alloys Compd. {\bf 338}, 126
(2002).

\bibitem{LE02}  E. M. Levin {\it et al.}, Phys. Rev. B {\bf 65}, 214427
(2002).

\bibitem{SO03}  J. B. Sousa {\it et al.}, Phys. Rev. B {\bf 67}, 134416
(2003).

\bibitem{HE02}  S. H\'{e}bert {\it et al.}, J. Solid State Chem. {\bf 165},
6 (2002); S. H\'{e}bert {\it et al}., Solid State Commun. {\bf 122}, 335 (2002).

\bibitem{MAI02}  A. Maignan {\it et al.}, J. Phys. Cond. Matt. {\bf 14}, 11809 (2002).

\bibitem{MAH02}  R. Mahendiran {\it et al.}, Phys. Rev. Lett. {\bf 89},
286602 (2002).

\bibitem{HA03}  V. Hardy {\it et al.}, J. Magn. Magn. Mater. {\bf 264}, 183
(2003).

\bibitem{NIS}  Z. Nishiyama, in {\it Martensitic Transformation}, edited by
M. Fine {\it et al.}, (Academic, New York, 1978).

\bibitem{MAR}  {\it International Conference on Martensitic Transformations}
(ICOMAT 95), Lausanne, 1995, edited by R. Gotthardt and J. Van Humbeeck,
(Editions de Physique, Les Ulis, 1995).

\bibitem{PO01}  V. Podzorov {\it et al.}, Phys. Rev. B {\bf 64}, 140406
(2001).

\bibitem{PER01}  F. J. Perez-Reche {\it et al.}, Phys. Rev. Lett. {\bf 87},
195701 (2001).
\end{thebibliography}
\end{document}